\documentclass[ aip,rsi,
 amsmath,amssymb,reprint,]{revtex4-1}

\usepackage{xcolor}
\usepackage{gensymb}
\usepackage{graphicx}
\usepackage{dcolumn}
\usepackage{bm}

\usepackage[utf8]{inputenc}
\usepackage[T1]{fontenc}
\usepackage{mathptmx}

\newcommand{\blu}{ }

\graphicspath{{Figs/}}
\begin{document}

\preprint{AIP/123-QED}

\title[ULF-NMR Coil System]{Fast switching coil system for sample premagnetization in an unshielded ultra-low-field \blu{Nuclear Magnetic Resonance experiment}}

\looseness=-1
\author{Valerio Biancalana}%
\email{\vspace{-8mm} valerio.biancalana@unisi.it}

\affiliation{Dept.\ of Information Engineering and Mathematics -  DIISM University of Siena - Via Roma 56, 53100 Siena, Italy \looseness=-1}
\author{Roberto Cecchi}
\author{Leonardo Stiaccini}
\affiliation{
Dept.\ of Physical Sciences, Earth and Environment - DSFTA University of Siena - Via Roma 56, 53100 Siena, Italy \looseness=-1
}

\author{Antonio Vigilante}
\affiliation{Dept.\ of Information Engineering and Mathematics -  DIISM University of Siena - Via Roma 56, 53100 Siena, Italy \looseness=-1}


\begin{abstract}
We present a system developed to premagnetize liquid samples in an ultra-low-field \blu{Nuclear Magnetic Resonance (NMR)} experiment. Liquid samples of a few milliliter are exposed to a magnetic field of about 70~mT, which is abruptely switched off, so to leave a transverse microtesla field where nuclei start precessing. An accurate characterization of the transients and intermediate field level enables a reliable operation of the detection system, which is based on an optical magnetometer. 
\vspace{-10mm}
\end{abstract}

\maketitle

The ultra-low-field \blu{(ULF)} regime in \blu{Nuclear Magnetic Resonance (NMR) experiments \cite{Kraus_2014}} corresponds to precession field intensities ranging around the microtesla level, where the nuclear precession is so slow to make the usual detection approaches --based on Faraday induction phenomenon-- inadequate or unfeasible. Instead, non-inductive, highly sensitive detectors can be used, such as SQUIDs and optical atomic magnetometers \cite{budker_nat_07}.

Optical magnetometers are acknowledged for their practicality (they operate at room temperature) and robustness, making them eligible as high-performance sensors in on-field applications, such as in tomography in hostile environment \cite{deans_ao_18} and material characterization \cite{bevington_apl_18} .

Optical magnetometers designed to operate in a weak field, can tolerate 
much stronger (e.g. tesla level) ones, and recover immediately their operativity, as soon as the weak (e.g. microtesla) field is restored. In ultra-low-field NMR, this is the case when the  sample is premagnetized in the same position where the nuclear precession is detected. Such \textit{in situ} magnetization constitutes an approach complementary to remote-detection techniques, where the sample is premagnetized by a strong field in a region at some distance from the sensors and is subsequently shuttled to the weak field region, in the proximity of the sensor that detects the nuclear precession.

This note describes an arrangement that makes possible to perform \textit{in situ} magnetization in a setup previously developed to register NMR \cite{biancalana_DH_jpcl_17, biancalana_zulfJcoupling_jmr_16} and MRI \cite{biancalana_IDEA_prappl_19, biancalana_apl_19} signals with a remote-detection scheme.

In most of \blu{\textit{in-situ}} ULF-NMR experimental setups, \textit{in-situ} premagnetization can be performed only with hyperpolarization techniques \cite{bowers_prl_86, reineri_cmra_06, eills_jacs_19}, due to incompatibility of magnetic shields with strong magnetic fields, so that Zeeman-interaction polarization with fields exceeding 50-100~mT is relegated to \textit{ex-situ} (remote detection) apparatuses \cite{tayler_rsi_17}. This is not the case of our setup, which is designed and optimized to operate in unshielded environment, and uses  active field-compensation \cite{biancalana_FPGAstab_prappl_19}  and  differential-measurement techniques \cite{biancalana_apb_16}   to reject disturbances produced by external sources, while other experiments in shielded volumes use typically weaker (10-20 mT) premagnetization fields \cite{tayler_apl_19}.

The detection geometry considered in this work is the same described in refs.\cite{biancalana_DH_jpcl_17, biancalana_zulfJcoupling_jmr_16}, but the NMR sample (about 7 milliliter of distilled water) is premagnetized by a room-temperature solenoidal coil placed in the proximity of the sensor. The coil is made of 500~g of 0.56~mm diameter copper wire, wound on its original reel. The coil assembly size and its electromagnetic characteristics are summarized in Table \ref{tab:coilparameters}.

\begin{table}[ht]
\vspace{-2mm}
\centering
\begin{tabular}{ | m{4.9cm} | m{20mm} |  } 
\hline
 \bf parameter & \bf value  \\ 
\hline
 Solenoid length & 54 mm \\ 
 \hline
  Wire length &  220 m \\ 
 \hline
  Number of turns &  2000 \\ 
 \hline
 Coil external diameter &  50 mm \\ 
 \hline
Resistance \blu{($R$)}, Inductance \blu{($L$)}&  15 $\Omega$, 64 mH \\ 
 \hline
Field & 40 mT/A  \\
\hline
\end{tabular}
\caption{Main features of the premagnetizing coil.}
\label{tab:coilparameters}
\vspace{-4mm}
\end{table}

The coil is inserted in a sealed PVC cylindrical shell, where a coolant liquid flows. The external radius of the shell (35 mm) summed with the the external radius of a cylinder containing the atomic sensor and its electric heater (40~mm), establishes a constraint that limits the minimum sample-sensor distance: a critical parameter setting the sample-sensor coupling, and eventually the signal detection efficiency.

A challenging aspect of the described setup concerns the need of reliable control of diverse magnetic field strength, with the twofold need of producing rather strong field and to generate fast, tailored field transitions. Driving inductive loads finds interests in a variety of industrial and research applications, where diverse kinds  of specifications and features are required, with parameters (such as peak current, power dissipation, compliance voltages, pulse duration etc.) ranging in broad intervals. Correspondingly, abundant literature can be found both among application notes of device producers \cite{long_maxim_17, forbes_tian_19, hopkins_st_12} and in scientific journals \cite{flaxer_rsi_01, guiji_rsi_13}. \blu{Particularly, fast transients on magnetic field are required in experiments involving cold atoms, degenerate gases and condensates \cite{aubin_jltp_05, herrera_phd_12}}.

The NMR signal detection is performed on the basis of a relatively simple sequence. The measurement is made in $B_m=4\, \mu$T field oriented along $y$. A much stronger (70~mT) $B_0$ field is  superimposed to $B_m$ in a transverse direction and premagnetizes the nuclear sample along $x$ for a few seconds (several times the nuclear longitudinal relaxation time $T_1$). This premagnetization field $B_0$ is reduced to a much weaker intensity $B_1$ (at least 10 times $B_m$), in a time as short as possible, compatibly with the current driver electric specifications (see below). As $\vec B_1$ is parallel to $\vec B_0$ and exceeds $B_m$ by at least an order of magnitude, the total field remains substantially parallel to the induced magnetization ($x$ direction). At this point the $B_1$ is abruptly turned to zero, in a time interval much shorter than the nuclear precession period. Due to such non-adiabatic field transition, the nuclear magnetization remains transverse to the field $B_m$ and starts precessing around the $y$ direction, at a frequency $\nu_n=\gamma_n B_m$ (170 Hz in the case of protons at $4\, \mu$T).

When the field has been restored at its detection level $B_m$, \blu{the magnetometer requires 
$25$ ms to recover its operativity, and other $25$ ms necessary to let the field-compensation system achieve its steady state operation.} After this dead-time, a time-dependent magnetic signal produced by the nuclear precession can be  detected and recorded. Repeated measurements are performed to improve the signal-to-noise ratio.

\begin{figure}[ht]
\vspace{-3mm}
    \centering
    \includegraphics [angle=0, width=0.84\columnwidth]{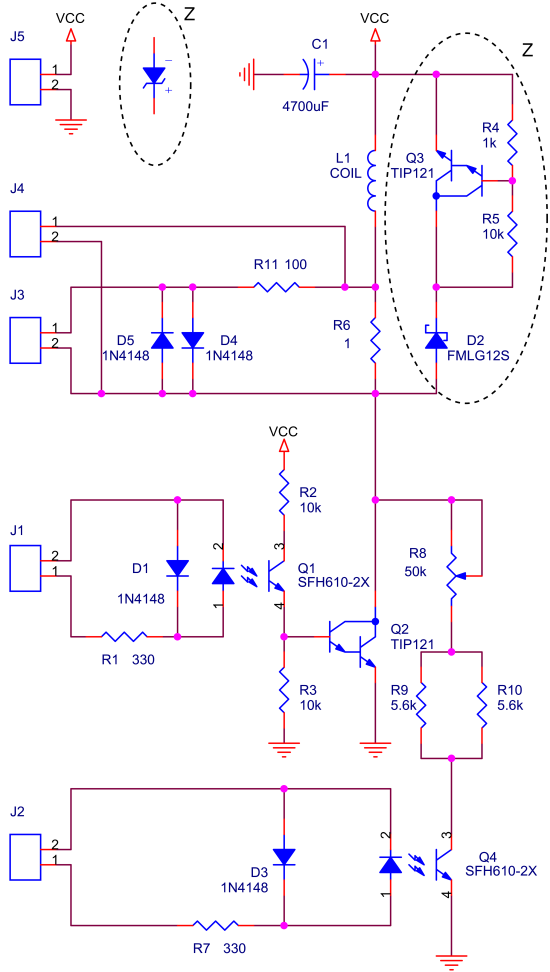}
    \caption{Schematics of the coil current driver. The coil supply is controlled by two transistors: Q2 is a Darlington Transistor --controlled by the opto-coupler Q1-- and drives the high-level current, the opto-coupler Q4 drives the low-level current, which is switched off with some delay after the high-level one. The extra-voltage produced by high-level current switch-off is limited to a non-destructive level by the branch made of D2, R4, R5, Q3. The latter branch behaves as a Zener diode Z (highlighted in dashed-black). The high-power series resistor R6 ($1\, \Omega$, 10~W) produces an output voltage that can be used as it is (on J4) for high-current monitor or  --after the clipping circuit made of R11, D4, D5-- (on J3) to monitor the low-current dynamics.  }
\label{fig:schema}
\vspace{-2mm}
\end{figure}

The coil producing the premagnetization field is supplied by a circuit which provides relatively high current (for nuclear spin premagnetization) and guarantees adjustable and precise current level transitions (for nuclear spin manipulation).

The current in the coil is controlled by two digital signals coming from a DAC board connected to the driver by opto-couplers. The high-current control signal is amplified by Q1-Q2, while the low-current is directly controlled by the opto-coupler Q4 (see Fig.\ref{fig:schema}). In the considered application, the onset dynamics of the current is irrelevant, while the switch-off requires to be fast and appropriately designed.

In order to dissipate the energy stored in the coil in a time shorter than the coil L/R characteristic time,  a parallel branch comes into play during the high-current switching off (at the high-to-low transition of the signal in J1), i.e. when $B_0$ is reduced to $B_1\approx 10\ B_m$. The reverse extra-voltage across the load is maintained below a given value $V_Z$ as to prevent that the  collector-emitter voltage $V_{CE}$ on the transistor Q2 to exceed the maximum rating ($V_{CE\text{,Max}}=$80~V for the selected device, TIP121). 

The time-derivative $dI/dt$ of the coil current is higher the higher is the voltage $V_Z$ in the Zener-like parallel circuit branch made of D2, R4, R5, Q3: 
 \vspace{-2mm}
\begin{equation}
    \frac{dI}{dt} \approx -\frac{1}{L}\left(V_Z+RI\right),
 \vspace{-2mm}
\end{equation}
where L and R are the inductance and the resistance of the load. Thus the speed of the $I_0\rightarrow I_1$ transition is enhanced for high $V_Z$, whose limit is set by the the above mentioned need of maintaining the $V_{CE}$ in Q2 at a safe level.

Once the coil reaches the steady state at $I_1$, a transition to zero of the signal in connector J2 controls the complete switching off of the coil current. This event causes the total field passing from $\vec B_1$ to $\vec B_m$, which is nearly perpendicular and at least ten times smaller. This transition lasts $T_t$, a time that must be shorter than the instantaneous nuclear precession period: $ T_t\leq2\pi/(\gamma_n B(t)) $ . This condition is easy fulfilled, even without activation of the parallel branch, provided that the $I_1$ level is adequately low.

During long-lasting operation the the coil temperature must be monitored to prevent over-heating. To this purpose the coil is packed and cooled using a water cooling system. \blu{This solution prevent over-heating when the coil is powered with a duty cycle of $\approx 40\%$. In addition} the current flowing in the coil is monitored acquiring the voltage across a 1$\, \Omega$ series resistance (R6 in Fig.\ref{fig:schema}). This signal can be used both to verify the actual time behaviour of the high-level and low-level current, and to estimate the current temperature during cycled measurements. 

As known, the copper resistivity increases by about 0.393\% per $\degree$C. As the circuit is supplied with a constant voltage, a reduction of 15\% in the maximum coil current indicates an average copper temperature increase of about 60$\degree$C. In this eventuality, the cycled data acquisition is suspended and the coil resistance is periodically measured, to restart the measurements when the coil is cooled down. 

A clipping circuit limits the monitor signal excursion and helps to prevent DAQ saturation when recording the low-level current dynamics.

\begin{figure}[ht]
\vspace{-4mm}
    \centering
    \includegraphics [angle=0, width=.95 \columnwidth]{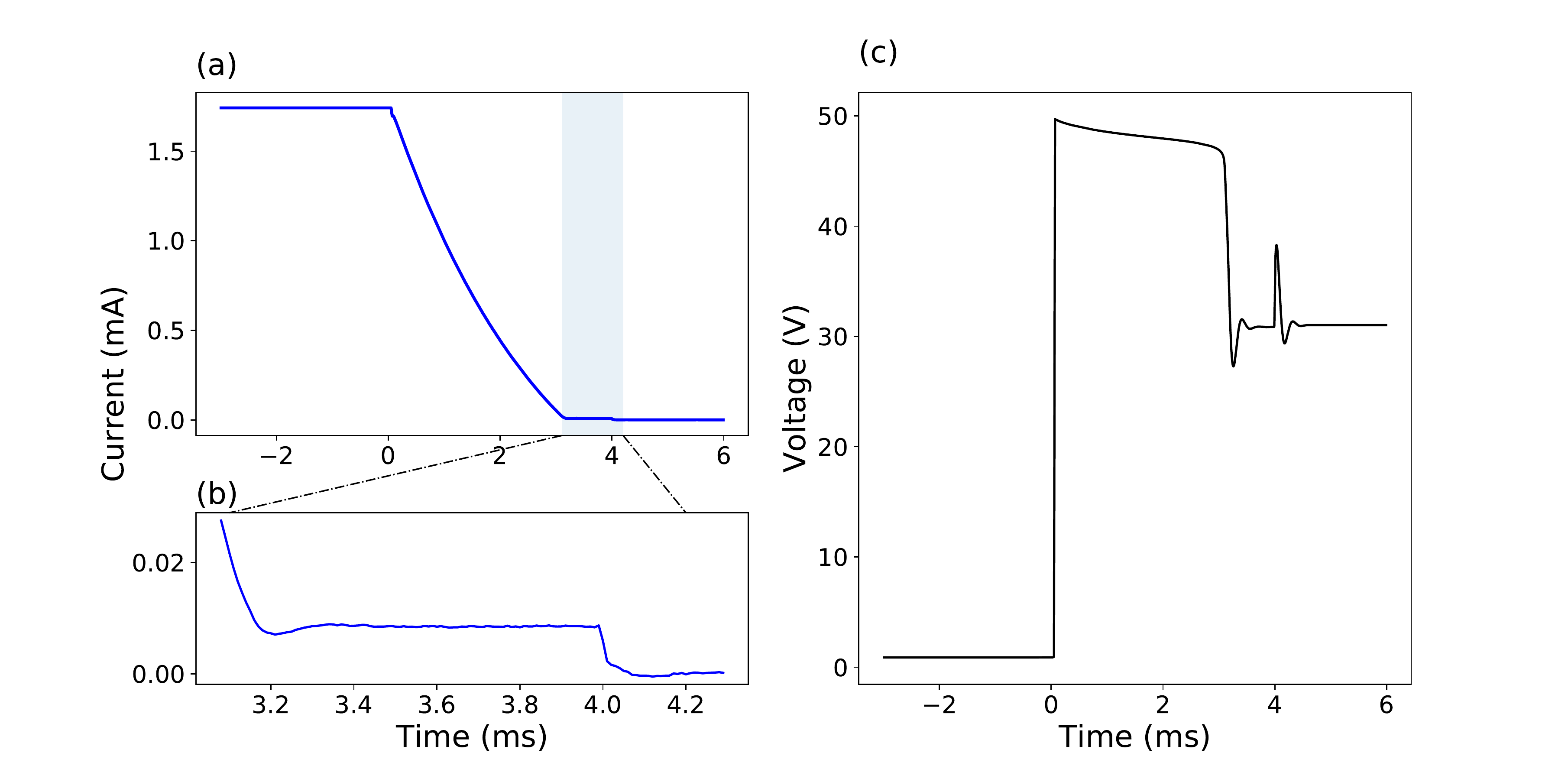}
    \vspace{-4mm}
    \caption{High-level (a) and low-level (b) coil current, and Q2 collector-emitter voltage (c) as a function of time, during the current switch-off. }
\label{fig:corrente}
\vspace{-4mm}
\end{figure}

The high-level and low-level current dynamics, are shown in Figs.\ref{fig:corrente} (a) and (b), while Fig.\ref{fig:corrente} (c) shows the $V_{CE}$ voltage on Q2. Fig.\ref{fig:corrente} (a) shows the transition from $I_0$ to $I_1$ with its constant slope in the time interval from 0 to 3~ms. 
The current reaches the $I_1$ level in about 3~ms with an appreciable but irrelevant overshot. The initial slope $dI/dt \approx -800\, $A/s on the load inductance $L=64\, $mH is consistent with the measured Q2 collector voltage $V_{CC}-L\ dI/dt -RI_0\approx 50$~V presented in Fig.\ref{fig:corrente} (c).

As shown in Fig.\ref{fig:corrente} (b), when (at t=4~ms) the J2 control voltage is set to zero, the   $I_1$ -to-zero transition occurs in about 10~$\mu$s. This second transition causes only a 10~V overvoltage peak, not enough to trigger the Zener protection. 

\begin{figure}[ht]
\vspace{-5mm}
    \centering
    \includegraphics [angle=0, width=.95 \columnwidth]{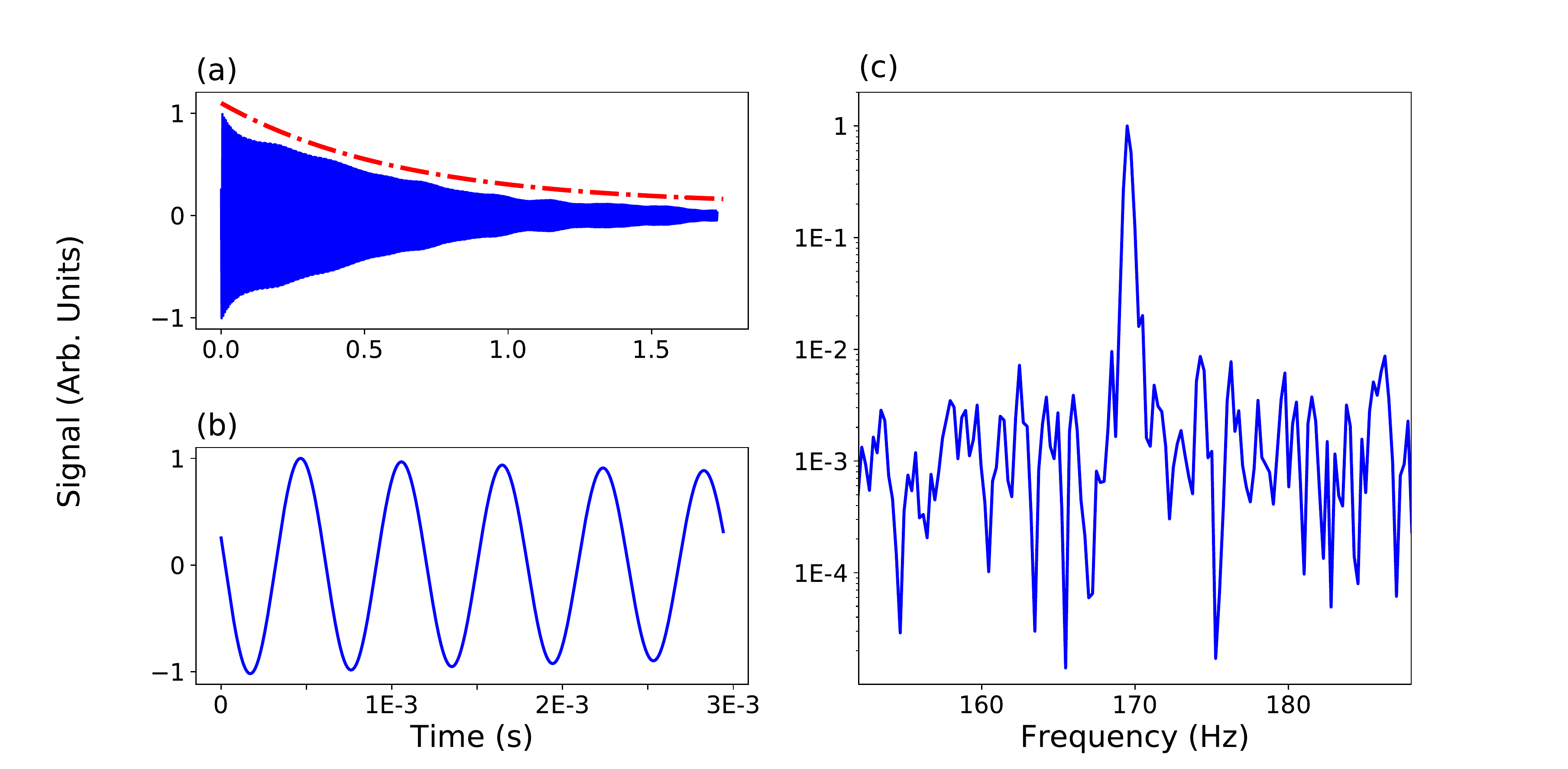}
    \vspace{-4mm}
    \caption{The ultra-low-field NMR signal resulting from a long lasting cycled measurement (about 1300 cycles). The signal in panel (a) is band-pass filtered around the proton NMR frequency  in a $4\ \mu$T field ($\approx$ 170 Hz). The panel (b) shows a small portion of that signal with an expanded time-scale, and the panel (c) shows the power spectrum of the raw signal\blu{, the figure shows a SNR of the order of $10^2$.The time signal (a) decays in $\approx 600$~ms decay time, consistently with the linewidth seen in the frequency domain (c).}}
\label{fig:NMR}
\vspace{-4mm}
\end{figure}

The described system has been designed to enable \textit{in situ} premagnetization of ultra-low-field NMR samples. Preliminary tests demonstrate the feasibility of such approach (thanks to the operation in unshielded environment).
The Fig.\ref{fig:NMR} shows the proton signal obtained at $4\ \mu$T precession field after premagnetization in 70~mT field followed by sudden (non adiabatic) reduction and 90$\degree$ rotation of the field. 
The detected signal shows a $\approx 600$~ms decay time, definitely shorter than in previous --remote-detection-- instances \cite{biancalana_DH_jpcl_17}. This is likely to be attributed to residual precession field inhomogeneities due to ferromagnetic contamination of the coil assembly materials \cite{tayler_apl_19, biancalana_submitted_20}.
\blu{Having selected the circuit devices to minimize the dark current helps also in preserving the magnetometric noise floor.}

In conclusion, we have developed a premagnetization system for a NMR experiment enabling the \textit{in-situ} premagnetization of the sample, and providing the magnetic field transitions suited to obtained nuclear spins tipped with respect to the ultra-low-field, where their precession is detected. We have demonstrated that a proton signal at 170~Hz with an initial amplitude of about 3~pT can be recorded from a 7~ml water sample premagnetized at about 70~mT.

The system could be rearranged to insert the sensor inside the coil, so to remarkably improve the sample-sensor coupling factor. The latter in fact dramatically scales with the sensor-sample distance, which in the present configuration is constrained by the coil thickness,  a reason for which further increase of the coil size would not help to enhance the signal.

The transition from the low-level current to zero is measured to be about 10 $\mu$s: it is fast (non-adiabatic) not only with respect to the proton precession, but also for the atomic precession in the magnetometer. The latter feature opens also the new perspective of performing the measurement in a different regime, where atoms are prepared in a stretched state during the premagnetization time, and are subsequently probed in a free precession condition --having turned off the pumping radiation-- during the NMR signal detection.


\bibliography{NewCoil}

\begin{thebibliography}{23}%
\makeatletter
\providecommand \@ifxundefined [1]{%
 \@ifx{#1\undefined}
}%
\providecommand \@ifnum [1]{%
 \ifnum #1\expandafter \@firstoftwo
 \else \expandafter \@secondoftwo
 \fi
}%
\providecommand \@ifx [1]{%
 \ifx #1\expandafter \@firstoftwo
 \else \expandafter \@secondoftwo
 \fi
}%
\providecommand \natexlab [1]{#1}%
\providecommand \enquote  [1]{``#1''}%
\providecommand \bibnamefont  [1]{#1}%
\providecommand \bibfnamefont [1]{#1}%
\providecommand \citenamefont [1]{#1}%
\providecommand \href@noop [0]{\@secondoftwo}%
\providecommand \href [0]{\begingroup \@sanitize@url \@href}%
\providecommand \@href[1]{\@@startlink{#1}\@@href}%
\providecommand \@@href[1]{\endgroup#1\@@endlink}%
\providecommand \@sanitize@url [0]{\catcode `\\12\catcode `\$12\catcode
  `\&12\catcode `\#12\catcode `\^12\catcode `\_12\catcode `\%12\relax}%
\providecommand \@@startlink[1]{}%
\providecommand \@@endlink[0]{}%
\providecommand \url  [0]{\begingroup\@sanitize@url \@url }%
\providecommand \@url [1]{\endgroup\@href {#1}{\urlprefix }}%
\providecommand \urlprefix  [0]{URL }%
\providecommand \Eprint [0]{\href }%
\providecommand \doibase [0]{http://dx.doi.org/}%
\providecommand \selectlanguage [0]{\@gobble}%
\providecommand \bibinfo  [0]{\@secondoftwo}%
\providecommand \bibfield  [0]{\@secondoftwo}%
\providecommand \translation [1]{[#1]}%
\providecommand \BibitemOpen [0]{}%
\providecommand \bibitemStop [0]{}%
\providecommand \bibitemNoStop [0]{.\EOS\space}%
\providecommand \EOS [0]{\spacefactor3000\relax}%
\providecommand \BibitemShut  [1]{\csname bibitem#1\endcsname}%
\let\auto@bib@innerbib\@empty
\bibitem [{\citenamefont {Kraus}\ \emph {et~al.}(2014)\citenamefont {Kraus},
  \citenamefont {Espy}, \citenamefont {Magnelind},\ and\ \citenamefont
  {Volegov}}]{Kraus_2014}%
  \BibitemOpen
  \bibfield  {author} {\bibinfo {author} {\bibfnamefont {R.}~\bibnamefont
  {Kraus}}, \bibinfo {author} {\bibfnamefont {M.}~\bibnamefont {Espy}},
  \bibinfo {author} {\bibfnamefont {P.}~\bibnamefont {Magnelind}}, \ and\
  \bibinfo {author} {\bibfnamefont {P.}~\bibnamefont {Volegov}},\ }\href
  {https://oxfordmedicine.com/view/10.1093/med/9780199796434.001.0001/med-9780199796434}
  {\emph {\bibinfo {title} {{U}ltra-{L}ow {F}ield {N}uclear {M}agnetic
  {R}esonance, {A} {N}ew {MRI R}egime}}}\ (\bibinfo  {publisher} {Oxford
  University Press},\ \bibinfo {address} {Oxford, UK},\ \bibinfo {year}
  {2014})\BibitemShut {NoStop}%
\bibitem [{\citenamefont {Budker}\ and\ \citenamefont
  {Romalis}(2007)}]{budker_nat_07}%
  \BibitemOpen
  \bibfield  {author} {\bibinfo {author} {\bibfnamefont {D.}~\bibnamefont
  {Budker}}\ and\ \bibinfo {author} {\bibfnamefont {M.}~\bibnamefont
  {Romalis}},\ }\href@noop {} {\bibfield  {journal} {\bibinfo  {journal}
  {Nature Physics}\ }\textbf {\bibinfo {volume} {3}},\ \bibinfo {pages} {227}
  (\bibinfo {year} {2007})}\BibitemShut {NoStop}%
\bibitem [{\citenamefont {Deans}, \citenamefont {Marmugi},\ and\ \citenamefont
  {Renzoni}(2018)}]{deans_ao_18}%
  \BibitemOpen
  \bibfield  {author} {\bibinfo {author} {\bibfnamefont {C.}~\bibnamefont
  {Deans}}, \bibinfo {author} {\bibfnamefont {L.}~\bibnamefont {Marmugi}}, \
  and\ \bibinfo {author} {\bibfnamefont {F.}~\bibnamefont {Renzoni}},\ }\href
  {\doibase 10.1364/AO.57.002346} {\bibfield  {journal} {\bibinfo  {journal}
  {Appl. Opt.}\ }\textbf {\bibinfo {volume} {57}},\ \bibinfo {pages} {2346}
  (\bibinfo {year} {2018})}\BibitemShut {NoStop}%
\bibitem [{\citenamefont {Bevington}\ \emph {et~al.}(2018)\citenamefont
  {Bevington}, \citenamefont {Gartman}, \citenamefont {Chalupczak},
  \citenamefont {Deans}, \citenamefont {Marmugi},\ and\ \citenamefont
  {Renzoni}}]{bevington_apl_18}%
  \BibitemOpen
  \bibfield  {author} {\bibinfo {author} {\bibfnamefont {P.}~\bibnamefont
  {Bevington}}, \bibinfo {author} {\bibfnamefont {R.}~\bibnamefont {Gartman}},
  \bibinfo {author} {\bibfnamefont {W.}~\bibnamefont {Chalupczak}}, \bibinfo
  {author} {\bibfnamefont {C.}~\bibnamefont {Deans}}, \bibinfo {author}
  {\bibfnamefont {L.}~\bibnamefont {Marmugi}}, \ and\ \bibinfo {author}
  {\bibfnamefont {F.}~\bibnamefont {Renzoni}},\ }\href {\doibase
  10.1063/1.5042033} {\bibfield  {journal} {\bibinfo  {journal} {Applied
  Physics Letters}\ }\textbf {\bibinfo {volume} {113}},\ \bibinfo {pages}
  {063503} (\bibinfo {year} {2018})},\ \Eprint
  {http://arxiv.org/abs/https://doi.org/10.1063/1.5042033}
  {https://doi.org/10.1063/1.5042033} \BibitemShut {NoStop}%
\bibitem [{\citenamefont {Bevilacqua}\ \emph {et~al.}(2017)\citenamefont
  {Bevilacqua}, \citenamefont {Biancalana}, \citenamefont {Dancheva},
  \citenamefont {Vigilante}, \citenamefont {Donati},\ and\ \citenamefont
  {Rossi}}]{biancalana_DH_jpcl_17}%
  \BibitemOpen
  \bibfield  {author} {\bibinfo {author} {\bibfnamefont {G.}~\bibnamefont
  {Bevilacqua}}, \bibinfo {author} {\bibfnamefont {V.}~\bibnamefont
  {Biancalana}}, \bibinfo {author} {\bibfnamefont {Y.}~\bibnamefont
  {Dancheva}}, \bibinfo {author} {\bibfnamefont {A.}~\bibnamefont {Vigilante}},
  \bibinfo {author} {\bibfnamefont {A.}~\bibnamefont {Donati}}, \ and\ \bibinfo
  {author} {\bibfnamefont {C.}~\bibnamefont {Rossi}},\ }\href@noop {}
  {\bibfield  {journal} {\bibinfo  {journal} {The Journal of Physical Chemistry
  Letters}\ }\textbf {\bibinfo {volume} {8}},\ \bibinfo {pages} {6176}
  (\bibinfo {year} {2017})}\BibitemShut {NoStop}%
\bibitem [{\citenamefont {Bevilacqua}\ \emph
  {et~al.}(2016{\natexlab{a}})\citenamefont {Bevilacqua}, \citenamefont
  {Biancalana}, \citenamefont {Ben Amar~Baranga}, \citenamefont {Dancheva},\
  and\ \citenamefont {Rossi}}]{biancalana_zulfJcoupling_jmr_16}%
  \BibitemOpen
  \bibfield  {author} {\bibinfo {author} {\bibfnamefont {G.}~\bibnamefont
  {Bevilacqua}}, \bibinfo {author} {\bibfnamefont {V.}~\bibnamefont
  {Biancalana}}, \bibinfo {author} {\bibfnamefont {A.}~\bibnamefont {Ben
  Amar~Baranga}}, \bibinfo {author} {\bibfnamefont {Y.}~\bibnamefont
  {Dancheva}}, \ and\ \bibinfo {author} {\bibfnamefont {C.}~\bibnamefont
  {Rossi}},\ }\href@noop {} {\bibfield  {journal} {\bibinfo  {journal} {Journal
  of Magnetic Resonance}\ }\textbf {\bibinfo {volume} {263}},\ \bibinfo {pages}
  {65} (\bibinfo {year} {2016}{\natexlab{a}})}\BibitemShut {NoStop}%
\bibitem [{\citenamefont {Bevilacqua}\ \emph
  {et~al.}(2019{\natexlab{a}})\citenamefont {Bevilacqua}, \citenamefont
  {Biancalana}, \citenamefont {Dancheva},\ and\ \citenamefont
  {Vigilante}}]{biancalana_IDEA_prappl_19}%
  \BibitemOpen
  \bibfield  {author} {\bibinfo {author} {\bibfnamefont {G.}~\bibnamefont
  {Bevilacqua}}, \bibinfo {author} {\bibfnamefont {V.}~\bibnamefont
  {Biancalana}}, \bibinfo {author} {\bibfnamefont {Y.}~\bibnamefont
  {Dancheva}}, \ and\ \bibinfo {author} {\bibfnamefont {A.}~\bibnamefont
  {Vigilante}},\ }\href@noop {} {\bibfield  {journal} {\bibinfo  {journal}
  {Phys. Rev. Applied}\ }\textbf {\bibinfo {volume} {11}},\ \bibinfo {pages}
  {024049} (\bibinfo {year} {2019}{\natexlab{a}})}\BibitemShut {NoStop}%
\bibitem [{\citenamefont {Bevilacqua}\ \emph
  {et~al.}(2019{\natexlab{b}})\citenamefont {Bevilacqua}, \citenamefont
  {Biancalana}, \citenamefont {Dancheva},\ and\ \citenamefont
  {Vigilante}}]{biancalana_apl_19}%
  \BibitemOpen
  \bibfield  {author} {\bibinfo {author} {\bibfnamefont {G.}~\bibnamefont
  {Bevilacqua}}, \bibinfo {author} {\bibfnamefont {V.}~\bibnamefont
  {Biancalana}}, \bibinfo {author} {\bibfnamefont {Y.}~\bibnamefont
  {Dancheva}}, \ and\ \bibinfo {author} {\bibfnamefont {A.}~\bibnamefont
  {Vigilante}},\ }\href@noop {} {\bibfield  {journal} {\bibinfo  {journal}
  {Applied Physics Letters}\ }\textbf {\bibinfo {volume} {115}},\ \bibinfo
  {pages} {174102} (\bibinfo {year} {2019}{\natexlab{b}})}\BibitemShut
  {NoStop}%
\bibitem [{\citenamefont {Bowers}\ and\ \citenamefont
  {Weitekamp}(1986)}]{bowers_prl_86}%
  \BibitemOpen
  \bibfield  {author} {\bibinfo {author} {\bibfnamefont {C.~R.}\ \bibnamefont
  {Bowers}}\ and\ \bibinfo {author} {\bibfnamefont {D.~P.}\ \bibnamefont
  {Weitekamp}},\ }\href {\doibase 10.1103/PhysRevLett.57.2645} {\bibfield
  {journal} {\bibinfo  {journal} {Phys. Rev. Lett.}\ }\textbf {\bibinfo
  {volume} {57}},\ \bibinfo {pages} {2645} (\bibinfo {year}
  {1986})}\BibitemShut {NoStop}%
\bibitem [{\citenamefont {Canet}\ \emph {et~al.}(2006)\citenamefont {Canet},
  \citenamefont {Aroulanda}, \citenamefont {Mutzenhardt}, \citenamefont {Aime},
  \citenamefont {Gobetto},\ and\ \citenamefont {Reineri}}]{reineri_cmra_06}%
  \BibitemOpen
  \bibfield  {author} {\bibinfo {author} {\bibfnamefont {D.}~\bibnamefont
  {Canet}}, \bibinfo {author} {\bibfnamefont {C.}~\bibnamefont {Aroulanda}},
  \bibinfo {author} {\bibfnamefont {P.}~\bibnamefont {Mutzenhardt}}, \bibinfo
  {author} {\bibfnamefont {S.}~\bibnamefont {Aime}}, \bibinfo {author}
  {\bibfnamefont {R.}~\bibnamefont {Gobetto}}, \ and\ \bibinfo {author}
  {\bibfnamefont {F.}~\bibnamefont {Reineri}},\ }\href {\doibase
  10.1002/cmr.a.20065} {\bibfield  {journal} {\bibinfo  {journal} {Concepts in
  Magnetic Resonance Part A}\ }\textbf {\bibinfo {volume} {28A}},\ \bibinfo
  {pages} {321} (\bibinfo {year} {2006})},\ \Eprint
  {http://arxiv.org/abs/https://onlinelibrary.wiley.com/doi/pdf/10.1002/cmr.a.20065}
  {https://onlinelibrary.wiley.com/doi/pdf/10.1002/cmr.a.20065} \BibitemShut
  {NoStop}%
\bibitem [{\citenamefont {Eills}\ \emph {et~al.}(2019)\citenamefont {Eills},
  \citenamefont {Cavallari}, \citenamefont {Carrera}, \citenamefont {Budker},
  \citenamefont {Aime},\ and\ \citenamefont {Reineri}}]{eills_jacs_19}%
  \BibitemOpen
  \bibfield  {author} {\bibinfo {author} {\bibfnamefont {J.}~\bibnamefont
  {Eills}}, \bibinfo {author} {\bibfnamefont {E.}~\bibnamefont {Cavallari}},
  \bibinfo {author} {\bibfnamefont {C.}~\bibnamefont {Carrera}}, \bibinfo
  {author} {\bibfnamefont {D.}~\bibnamefont {Budker}}, \bibinfo {author}
  {\bibfnamefont {S.}~\bibnamefont {Aime}}, \ and\ \bibinfo {author}
  {\bibfnamefont {F.}~\bibnamefont {Reineri}},\ }\href {\doibase
  10.1021/jacs.9b10094} {\bibfield  {journal} {\bibinfo  {journal} {Journal of
  the American Chemical Society}\ }\textbf {\bibinfo {volume} {141}},\ \bibinfo
  {pages} {20209} (\bibinfo {year} {2019})},\ \bibinfo {note} {pMID:
  31762271},\ \Eprint
  {http://arxiv.org/abs/https://doi.org/10.1021/jacs.9b10094}
  {https://doi.org/10.1021/jacs.9b10094} \BibitemShut {NoStop}%
\bibitem [{\citenamefont {Tayler}\ \emph {et~al.}(2017)\citenamefont {Tayler},
  \citenamefont {Theis}, \citenamefont {Sjolander}, \citenamefont {Blanchard},
  \citenamefont {Kentner}, \citenamefont {Pustelny}, \citenamefont {Pines},\
  and\ \citenamefont {Budker}}]{tayler_rsi_17}%
  \BibitemOpen
  \bibfield  {author} {\bibinfo {author} {\bibfnamefont {M.~C.~D.}\
  \bibnamefont {Tayler}}, \bibinfo {author} {\bibfnamefont {T.}~\bibnamefont
  {Theis}}, \bibinfo {author} {\bibfnamefont {T.~F.}\ \bibnamefont
  {Sjolander}}, \bibinfo {author} {\bibfnamefont {J.~W.}\ \bibnamefont
  {Blanchard}}, \bibinfo {author} {\bibfnamefont {A.}~\bibnamefont {Kentner}},
  \bibinfo {author} {\bibfnamefont {S.}~\bibnamefont {Pustelny}}, \bibinfo
  {author} {\bibfnamefont {A.}~\bibnamefont {Pines}}, \ and\ \bibinfo {author}
  {\bibfnamefont {D.}~\bibnamefont {Budker}},\ }\href@noop {} {\bibfield
  {journal} {\bibinfo  {journal} {Review of Scientific Instruments}\ }\textbf
  {\bibinfo {volume} {88}},\ \bibinfo {pages} {091101} (\bibinfo {year}
  {2017})}\BibitemShut {NoStop}%
\bibitem [{\citenamefont {Bevilacqua}\ \emph
  {et~al.}(2019{\natexlab{c}})\citenamefont {Bevilacqua}, \citenamefont
  {Biancalana}, \citenamefont {Dancheva},\ and\ \citenamefont
  {Vigilante}}]{biancalana_FPGAstab_prappl_19}%
  \BibitemOpen
  \bibfield  {author} {\bibinfo {author} {\bibfnamefont {G.}~\bibnamefont
  {Bevilacqua}}, \bibinfo {author} {\bibfnamefont {V.}~\bibnamefont
  {Biancalana}}, \bibinfo {author} {\bibfnamefont {Y.}~\bibnamefont
  {Dancheva}}, \ and\ \bibinfo {author} {\bibfnamefont {A.}~\bibnamefont
  {Vigilante}},\ }\href@noop {} {\bibfield  {journal} {\bibinfo  {journal}
  {Phys. Rev. Applied}\ }\textbf {\bibinfo {volume} {11}},\ \bibinfo {pages}
  {014029} (\bibinfo {year} {2019}{\natexlab{c}})}\BibitemShut {NoStop}%
\bibitem [{\citenamefont {Bevilacqua}\ \emph
  {et~al.}(2016{\natexlab{b}})\citenamefont {Bevilacqua}, \citenamefont
  {Biancalana}, \citenamefont {Chessa},\ and\ \citenamefont
  {Dancheva}}]{biancalana_apb_16}%
  \BibitemOpen
  \bibfield  {author} {\bibinfo {author} {\bibfnamefont {G.}~\bibnamefont
  {Bevilacqua}}, \bibinfo {author} {\bibfnamefont {V.}~\bibnamefont
  {Biancalana}}, \bibinfo {author} {\bibfnamefont {P.}~\bibnamefont {Chessa}},
  \ and\ \bibinfo {author} {\bibfnamefont {Y.}~\bibnamefont {Dancheva}},\
  }\href@noop {} {\bibfield  {journal} {\bibinfo  {journal} {Applied Physics
  B}\ }\textbf {\bibinfo {volume} {122}},\ \bibinfo {pages} {103} (\bibinfo
  {year} {2016}{\natexlab{b}})}\BibitemShut {NoStop}%
\bibitem [{\citenamefont {Tayler}, \citenamefont {Ward-Williams},\ and\
  \citenamefont {Gladden}(2019)}]{tayler_apl_19}%
  \BibitemOpen
  \bibfield  {author} {\bibinfo {author} {\bibfnamefont {M.~C.~D.}\
  \bibnamefont {Tayler}}, \bibinfo {author} {\bibfnamefont {J.}~\bibnamefont
  {Ward-Williams}}, \ and\ \bibinfo {author} {\bibfnamefont {L.~F.}\
  \bibnamefont {Gladden}},\ }\href {\doibase 10.1063/1.5110658} {\bibfield
  {journal} {\bibinfo  {journal} {Applied Physics Letters}\ }\textbf {\bibinfo
  {volume} {115}},\ \bibinfo {pages} {072409} (\bibinfo {year} {2019})},\
  \Eprint {http://arxiv.org/abs/https://doi.org/10.1063/1.5110658}
  {https://doi.org/10.1063/1.5110658} \BibitemShut {NoStop}%
\bibitem [{\citenamefont {Long}(2017)}]{long_maxim_17}%
  \BibitemOpen
  \bibfield  {author} {\bibinfo {author} {\bibfnamefont {S.}~\bibnamefont
  {Long}},\ }\href
  {https://www.maximintegrated.com/en/design/technical-documents/app-notes/6/6307.html}
  {\bibfield  {journal} {\bibinfo  {journal} {Maxim Integrated Application Note
  6307}\ } (\bibinfo {year} {2017})}\BibitemShut {NoStop}%
\bibitem [{\citenamefont {Forbes}\ \emph {et~al.}(2019)\citenamefont {Forbes},
  \citenamefont {Harmouch}, \citenamefont {Unnikrishnan},\ and\ \citenamefont
  {Phillips}}]{forbes_tian_19}%
  \BibitemOpen
  \bibfield  {author} {\bibinfo {author} {\bibfnamefont {A.}~\bibnamefont
  {Forbes}}, \bibinfo {author} {\bibfnamefont {M.}~\bibnamefont {Harmouch}},
  \bibinfo {author} {\bibfnamefont {S.}~\bibnamefont {Unnikrishnan}}, \ and\
  \bibinfo {author} {\bibfnamefont {C.}~\bibnamefont {Phillips}},\ }\href
  {http://www.ti.com/lit/an/slvae30a/slvae30a.pdf} {\bibfield  {journal}
  {\bibinfo  {journal} {Texas Instruments Application Note slvae30a}\ }
  (\bibinfo {year} {2019})}\BibitemShut {NoStop}%
\bibitem [{\citenamefont {Hopkins}(2013)}]{hopkins_st_12}%
  \BibitemOpen
  \bibfield  {author} {\bibinfo {author} {\bibfnamefont {T.}~\bibnamefont
  {Hopkins}},\ }\href
  {https://www.st.com/content/ccc/resource/technical/document/application_note/a8/8c/73/84/dd/b2/44/75/CD00003790.pdf/files/CD00003790.pdf/jcr:content/translations/en.CD00003790.pdf}
  {\bibfield  {journal} {\bibinfo  {journal} {STMicroelectronics Application
  Note 280}\ } (\bibinfo {year} {2013})}\BibitemShut {NoStop}%
\bibitem [{\citenamefont {Flaxer}(2002)}]{flaxer_rsi_01}%
  \BibitemOpen
  \bibfield  {author} {\bibinfo {author} {\bibfnamefont {E.}~\bibnamefont
  {Flaxer}},\ }\href {\doibase 10.1063/1.1469673} {\bibfield  {journal}
  {\bibinfo  {journal} {Review of Scientific Instruments}\ }\textbf {\bibinfo
  {volume} {73}},\ \bibinfo {pages} {2197} (\bibinfo {year} {2002})},\ \Eprint
  {http://arxiv.org/abs/https://doi.org/10.1063/1.1469673}
  {https://doi.org/10.1063/1.1469673} \BibitemShut {NoStop}%
\bibitem [{\citenamefont {Wang}\ \emph {et~al.}(2013)\citenamefont {Wang},
  \citenamefont {Luo}, \citenamefont {Zhang}, \citenamefont {Zhao},
  \citenamefont {Sun}, \citenamefont {Tan}, \citenamefont {Chong},
  \citenamefont {Mo}, \citenamefont {Wu},\ and\ \citenamefont
  {Tao}}]{guiji_rsi_13}%
  \BibitemOpen
  \bibfield  {author} {\bibinfo {author} {\bibfnamefont {G.}~\bibnamefont
  {Wang}}, \bibinfo {author} {\bibfnamefont {B.}~\bibnamefont {Luo}}, \bibinfo
  {author} {\bibfnamefont {X.}~\bibnamefont {Zhang}}, \bibinfo {author}
  {\bibfnamefont {J.}~\bibnamefont {Zhao}}, \bibinfo {author} {\bibfnamefont
  {C.}~\bibnamefont {Sun}}, \bibinfo {author} {\bibfnamefont {F.}~\bibnamefont
  {Tan}}, \bibinfo {author} {\bibfnamefont {T.}~\bibnamefont {Chong}}, \bibinfo
  {author} {\bibfnamefont {J.}~\bibnamefont {Mo}}, \bibinfo {author}
  {\bibfnamefont {G.}~\bibnamefont {Wu}}, \ and\ \bibinfo {author}
  {\bibfnamefont {Y.}~\bibnamefont {Tao}},\ }\href {\doibase 10.1063/1.4788935}
  {\bibfield  {journal} {\bibinfo  {journal} {Review of Scientific
  Instruments}\ }\textbf {\bibinfo {volume} {84}},\ \bibinfo {pages} {015117}
  (\bibinfo {year} {2013})},\ \Eprint
  {http://arxiv.org/abs/https://doi.org/10.1063/1.4788935}
  {https://doi.org/10.1063/1.4788935} \BibitemShut {NoStop}%
\bibitem [{\citenamefont {{Aubin}}\ \emph {et~al.}(2005)\citenamefont
  {{Aubin}}, \citenamefont {{Extavour}}, \citenamefont {{Myrskog}},
  \citenamefont {{Leblanc}}, \citenamefont {{Est{\`e}ve}}, \citenamefont
  {{Singh}}, \citenamefont {{Scrutton}}, \citenamefont {{McKay}}, \citenamefont
  {{McKenzie}}, \citenamefont {{Leroux}}, \citenamefont {{Stummer}},\ and\
  \citenamefont {{Thywissen}}}]{aubin_jltp_05}%
  \BibitemOpen
  \bibfield  {author} {\bibinfo {author} {\bibfnamefont {S.}~\bibnamefont
  {{Aubin}}}, \bibinfo {author} {\bibfnamefont {M.~H.~T.}\ \bibnamefont
  {{Extavour}}}, \bibinfo {author} {\bibfnamefont {S.}~\bibnamefont
  {{Myrskog}}}, \bibinfo {author} {\bibfnamefont {L.~J.}\ \bibnamefont
  {{Leblanc}}}, \bibinfo {author} {\bibfnamefont {J.}~\bibnamefont
  {{Est{\`e}ve}}}, \bibinfo {author} {\bibfnamefont {S.}~\bibnamefont
  {{Singh}}}, \bibinfo {author} {\bibfnamefont {P.}~\bibnamefont {{Scrutton}}},
  \bibinfo {author} {\bibfnamefont {D.}~\bibnamefont {{McKay}}}, \bibinfo
  {author} {\bibfnamefont {R.}~\bibnamefont {{McKenzie}}}, \bibinfo {author}
  {\bibfnamefont {I.~D.}\ \bibnamefont {{Leroux}}}, \bibinfo {author}
  {\bibfnamefont {A.}~\bibnamefont {{Stummer}}}, \ and\ \bibinfo {author}
  {\bibfnamefont {J.~H.}\ \bibnamefont {{Thywissen}}},\ }\href {\doibase
  10.1007/s10909-005-7322-5} {\bibfield  {journal} {\bibinfo  {journal}
  {Journal of Low Temperature Physics}\ }\textbf {\bibinfo {volume} {140}},\
  \bibinfo {pages} {377} (\bibinfo {year} {2005})},\ \Eprint
  {http://arxiv.org/abs/cond-mat/0502196} {arXiv:cond-mat/0502196
  [cond-mat.soft]} \BibitemShut {NoStop}%
\bibitem [{\citenamefont {Herrera~Benzaquén}(2012)}]{herrera_phd_12}%
  \BibitemOpen
  \bibfield  {author} {\bibinfo {author} {\bibfnamefont {I.}~\bibnamefont
  {Herrera~Benzaquén}},\ }\emph {\bibinfo {title} {Degenerate quantum gases
  production and coherent manipulation on atom chips}},\ \href@noop {}
  {\bibinfo {type} {Phd thesis}},\ \bibinfo  {school} {Quantum Gases" at LENS
  Institute of the University of Florence - Universidad Autónoma de Madrid}
  (\bibinfo {year} {2012})\BibitemShut {NoStop}%
\bibitem [{\citenamefont {Bevilacqua}\ \emph {et~al.}(2020)\citenamefont
  {Bevilacqua}, \citenamefont {Biancalana}, \citenamefont {Consumi},
  \citenamefont {Dancheva}, \citenamefont {Rossi}, \citenamefont {Stiaccini},\
  and\ \citenamefont {Vigilante}}]{biancalana_submitted_20}%
  \BibitemOpen
  \bibfield  {author} {\bibinfo {author} {\bibfnamefont {G.}~\bibnamefont
  {Bevilacqua}}, \bibinfo {author} {\bibfnamefont {V.}~\bibnamefont
  {Biancalana}}, \bibinfo {author} {\bibfnamefont {M.}~\bibnamefont {Consumi}},
  \bibinfo {author} {\bibfnamefont {Y.}~\bibnamefont {Dancheva}}, \bibinfo
  {author} {\bibfnamefont {C.}~\bibnamefont {Rossi}}, \bibinfo {author}
  {\bibfnamefont {L.}~\bibnamefont {Stiaccini}}, \ and\ \bibinfo {author}
  {\bibfnamefont {A.}~\bibnamefont {Vigilante}},\ }\href@noop {} {\enquote
  {\bibinfo {title} {Ferromagnetic contamination of ultra-low-field-nmr sample
  containers. quantification of the problem and possible solutions.}}\ }
  (\bibinfo {year} {2020}),\ \bibinfo {note} {submitted}\BibitemShut {NoStop}%
\end{thebibliography}%

\end{document}